\documentclass[final,1p,times]{elsarticle}

\pdfoutput=1
\journal{Nuclear Inst. and Meth. in Phys. Res., A, NIMA-D-14-00850}

\usepackage{amssymb}
\usepackage{a4wide}
\usepackage{calc}
\usepackage{fancyhdr}
\usepackage{graphicx}
\usepackage[english]{babel}
\usepackage[bookmarks]{hyperref}
\usepackage[version=3]{mhchem}
\usepackage{multirow}
\usepackage{lmodern}
\usepackage[usenames, dvipsnames]{color}
\usepackage{tabularx}
\usepackage{floatrow}
\usepackage{rotating}
\usepackage{longtable}
\usepackage{booktabs}
\usepackage{amssymb}
\usepackage{amsmath}
\usepackage{enumitem}
\usepackage{gensymb}
\usepackage{amsfonts}
\usepackage{pdflscape}
\usepackage{bm}
\usepackage[mathscr]{euscript}

\usepackage[%
    format=hang,    
    format=plain,
    margin=0pt,
    width=0.8\textwidth,
]{caption}
\usepackage[list=true,listformat=simple,labelformat=simple]{subcaption}%labelformat=simple
\usepackage{url}
\usepackage{hyperref}
\usepackage{epstopdf}
\begin{document}
%\linenumbers
\begin{frontmatter}

%% Title, authors and addresses

%% use the tnoteref command within \title for footnotes;
%% use the tnotetext command for theassociated footnote;
%% use the fnref command within \author or \address for footnotes;
%% use the fntext command for theassociated footnote;
%% use the corref command within \author for corresponding author footnotes;
%% use the cortext command for theassociated footnote;
%% use the ead command for the email address,
%% and the form \ead[url] for the home page:
%% \title{Title\tnoteref{label1}}
%% \tnotetext[label1]{}
%% \author{Name\corref{cor1}\fnref{label2}}
%% \ead{email address}
%% \ead[url]{home page}
%% \fntext[label2]{}
%% \cortext[cor1]{}
%% \address{Address\fnref{label3}}
%% \fntext[label3]{}

\title{Alpha-event and surface characterisation in segmented true-coaxial HPGe detectors}

%% use optional labels to link authors explicitly to addresses:
%% \author[label1,label2]{}
%% \address[label1]{}
%% \address[label2]{}

\author[a]{I.~Abt}
\author[a]{L.~Garbini\corref{cor1}}
\ead{luciagarbini86@gmail.com}
\author[a]{C.~Gooch}
\author[a]{S.~Irlbeck}
\author[a]{X.~Liu}
\author[a,deg2]{M.~Palermo}
\author[a]{O.~Schulz}
\address[a]{Max-Planck-Institut f\"{u}r Physik, Munich, Germany}
%\address[b]{Now at Physics and Astronomy Department, University of Hawaii atManoa, HI, US}
\cortext[cor1]{Corresponding Author}
\fntext[deg2]{Now at Physics and Astronomy Department, University of Hawaii at Manoa, HI, US}

\begin{abstract}

A detailed study of alpha interactions on the passivation layer on the
end-plate of a true-coaxial high-purity germanium detector is presented.
The observation of alpha events on such a surface indicates an
unexpectedly thin so-called ``effective dead layer'' of less 
than 20\,$\mu$m thickness.
In addition, the influence of the metalisation close to the end-plate
on the time evolution of the output pulses is discussed.
The results indicate that alpha contamination can result in
events which could be mistaken as signals for neutrinoless
double beta decay and provide some guidance on how to prevent this.
\end{abstract}


\end{frontmatter}

%\linenumbers

%% main text
\section{Introduction}\label{section:introduction}

The search for neutrinoless double-beta ($0\nu\beta\beta$) decay 
is one of the most
promising approaches to look for physics beyond the standard
model. The germanium isotope \ce{^{76}Ge} is one of the 
candidates for such a search.
The current generation of germanium experiments, GERDA~\cite{Abt:2004yk}
and the MAJORANA demonstrator~\cite{Abgrall:2013rze}, 
are already setting significant
limits on the half-life of \ce{^{76}Ge} for $0\nu\beta\beta$ decay~\cite{GERDA_res16, Elliott:2016ble}  
and are working hard to push the currently available technologies
to reduce the background.
Recently, the GERDA collaboration has presented limits on 
the half-life of $5.2 \times 10^{25}$\,y and
on the mass of the coherent Majorana state, $m_{ee}$, 
between 160 and 400\,meV, depending
on the nuclear matrix element used. A background index of $\approx 10^{-3}$\,counts/keV/kg/y was also presented.
A commonly accepted goal is to increase the sensitivity down to 10\,meV in
order to cover the parameter space associated with the inverted 
hierarchy~\cite{Feruglio:2002}. This requires a ton-scale experiment with
a background index of better than 
10$^{-4}$\,counts/keV/kg/y~\cite{Maneschg2015188}.

Similar considerations lead also to the desire to build a ton-scale
germanium experiment to search for dark
matter~\cite{Aalseth:2012if, Kang:2013sjq, Ruppin:2014bra}.
While true-coaxial detectors are not candidates for such experiments,
the relevant detectors still have passivated or oxydised
surfaces and alpha interactions are relevant.

The choice of detector technology for any ton-scale experiment
needs careful consideration. 
The study presented here provides some input to the decision making process.
Large coaxial detectors are candidates for experiments searching for
$0\nu\beta\beta$ decay because they can be built with masses of more
than 2\,kg.
However, they have at least one floating end-plate,
which is usually passivated. 
Any contamination of these end-plates with alpha emitters has to be
considered~\cite{Johnson:2012pq}.
Several hundred large detectors are required
to achieve the necessary mass and thus the detector technology should
be commercially available. 
Therefore, it is important to investigate detectors which
are based on such technology.

The detector used for the study presented here was manufactured 
by Canberra France, now Mirion technology.
It has a standard end-plate geometry, but a special segmentation
to facilitate the characterisation of the end-plate.
Data are presented for which the passivated top surface 
was scanned with a \ce{^{241}Am} source.
It is shown that alpha particles are observed, even though their
range in germanium is less than 30\,$\mu$m.
The events induced by the alpha particles are characterised. 
The events are influenced by a 
so-called ``inactive'' or
``dead layer'' underneath the  passivated surface.
The simplistic term ``dead layer'', which is commonly 
used~\cite{dinger:1975,utso:2005,eberth:2008,maggi:2015},
does not reflect what happens
inside the detector. 
The actual physics of this region is involved; this region is neither inactive
nor dead. 
Throughout this paper, the term ``effective dead layer''
is used. 
The thickness of this layer is derived by converting the
energy, which could not be observed, to the thickness of a layer
of germanium, which an alpha particle would have to traverse to lose
the same amount of energy.   
As this apparent thickness can at least be partly due to charge
trapping along the drift paths,
it can be different for different charge carriers.

The important question for a future large-scale experiment is
whether alpha particles entering from the top of the detector can be misidentified
as $0\nu\beta\beta$ events if only the energy observed through
the detector core is recorded.
In addition, the influence of the metalisation close to the end-plate
is important. As the aluminium of the metalisation could also be
a source of background~\cite{Majorovits201139}, its influence was 
also investigated.

\section{Experimental setup and detector}\label{sec:setup} 

\begin{figure}[!h]
 \centering
  \includegraphics[width=0.80\linewidth]{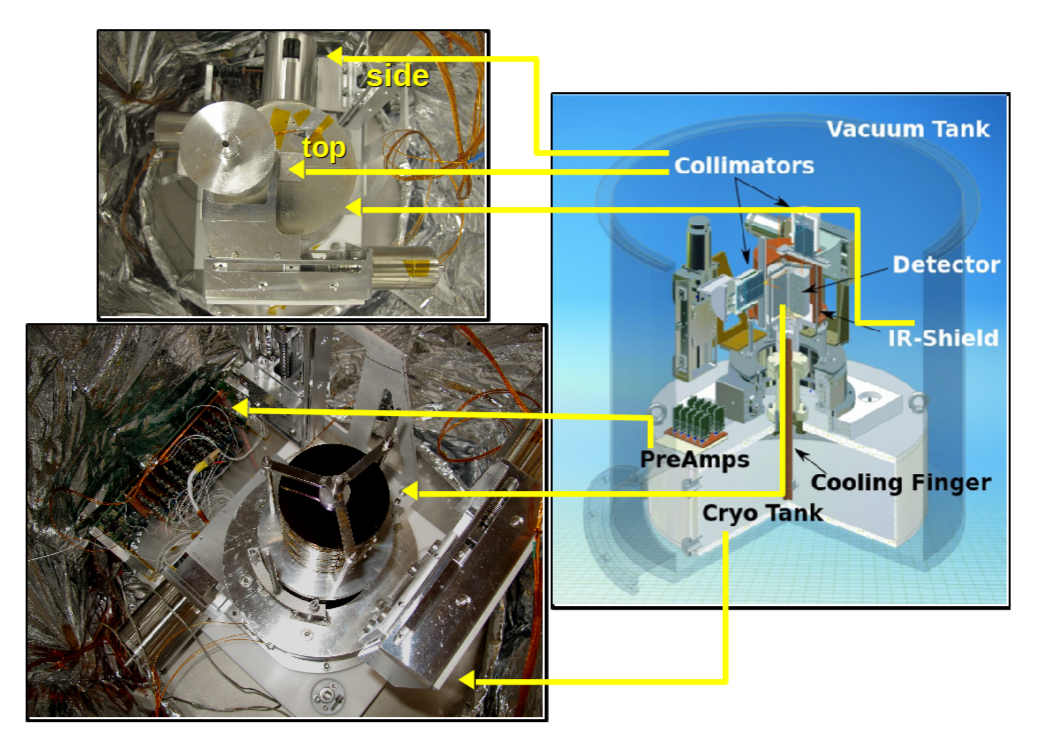}
\caption[Experimental setup]{Operation of the Super-Siegfried detector in the GALATEA test facility, adapted from~\cite{Garbini:2016}. }\label{fig:setup}
\end{figure}

The measurements presented in this paper were performed 
operating the ``Super-Siegfried'' test-detector in the 
GALATEA test-facility~\cite{Abt:2014bpa, Irlbeck:2010} 
shown in Fig.~\ref{fig:setup}, located at the Max-Planck-Institut 
f\"{u}r Physik in Munich. 
GALATEA is based on a vacuum chamber 
in which germanium detectors can be scanned 
by a system of three motorised stages.
Two tungsten collimators are used to focus the radiation onto 
the mantle and the top plate of any cylindrical detector.
As the paths between the collimators and the detector are free of material,
it is possible to use alpha sources.

\begin{figure}[!h]
 \centering
   \begin{subfigure}{.5\textwidth}
   \centering
   \includegraphics[width=.7\textwidth]{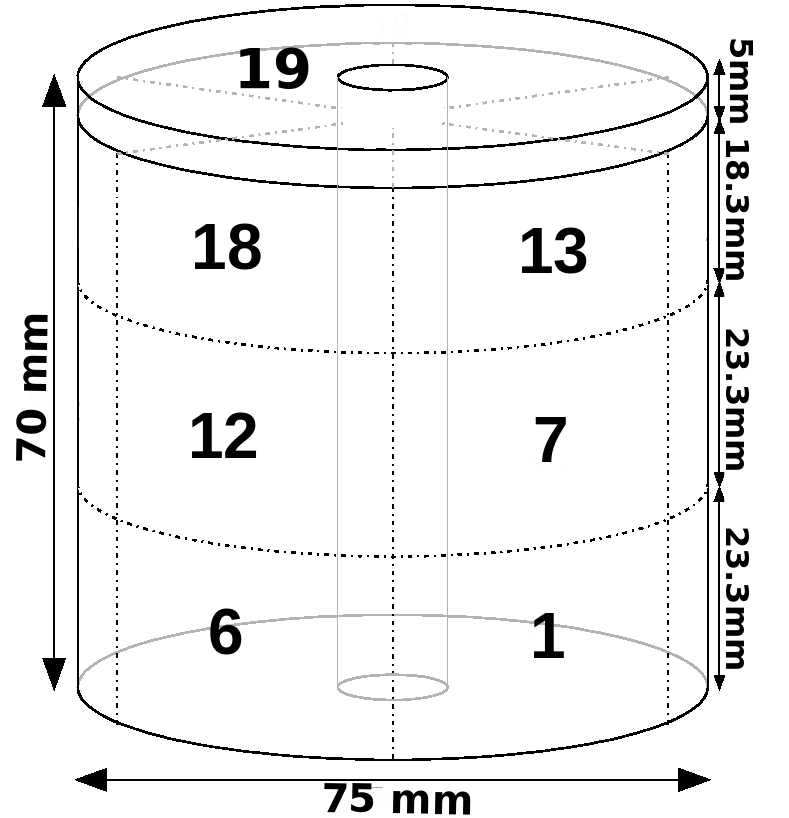}
   \caption[SuSie 3D frame]{\label{fig:susie3dScheme}}
   \end{subfigure}%
   \begin{subfigure}{.5\textwidth}
   \centering 
   \includegraphics[width=.9\textwidth]{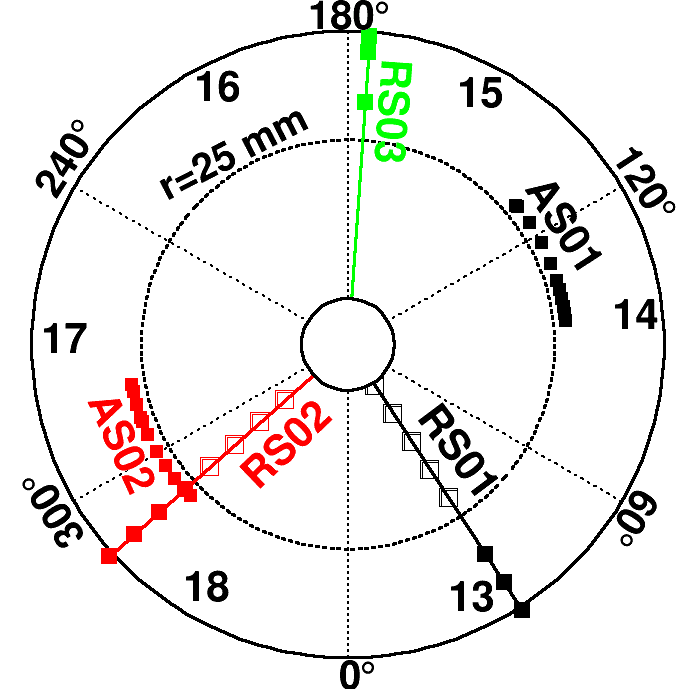}  
   \caption[Scanning Points]{\label{fig:scanning_points}}   
   \end{subfigure}
    \begin{subfigure}{.8\textwidth}
    \centering 
   \includegraphics[width=0.95\textwidth]{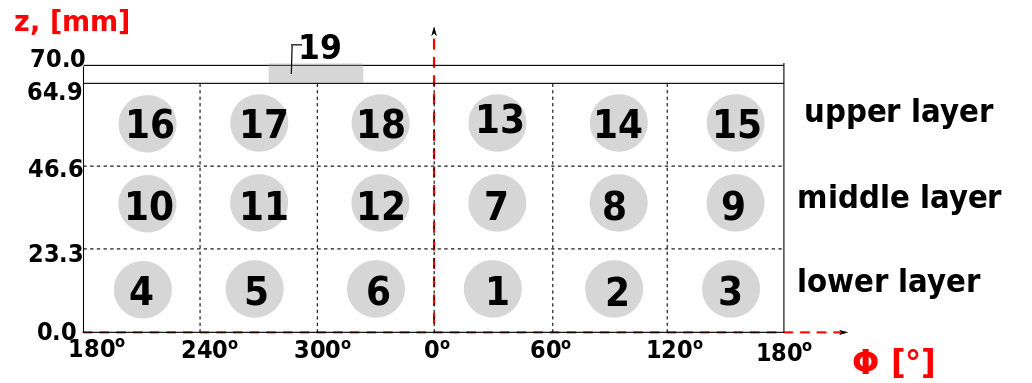}
    \caption[SuSie Open Frame]{\label{fig:susieOpenFrame}}
    \end{subfigure}%
\caption[Susie]{(a) Schematic of the 18+1\,fold segmented Super-Siegfried
      detector, (b) the locations of the individual measurements on the 
      top surface.
      The numbers indicate the segments underneath. Open symbols indicate no
      observation of alpha interaction at the probed location, (c) rolled out detector mantle with dimensions and reference system. The metalised contacts are indicated as shaded areas, adapted from~\cite{Garbini:2016}
}
\label{fig:Susie}
\end{figure}

\begin{figure}[!h]
 \centering
  \includegraphics[width=0.65\linewidth]{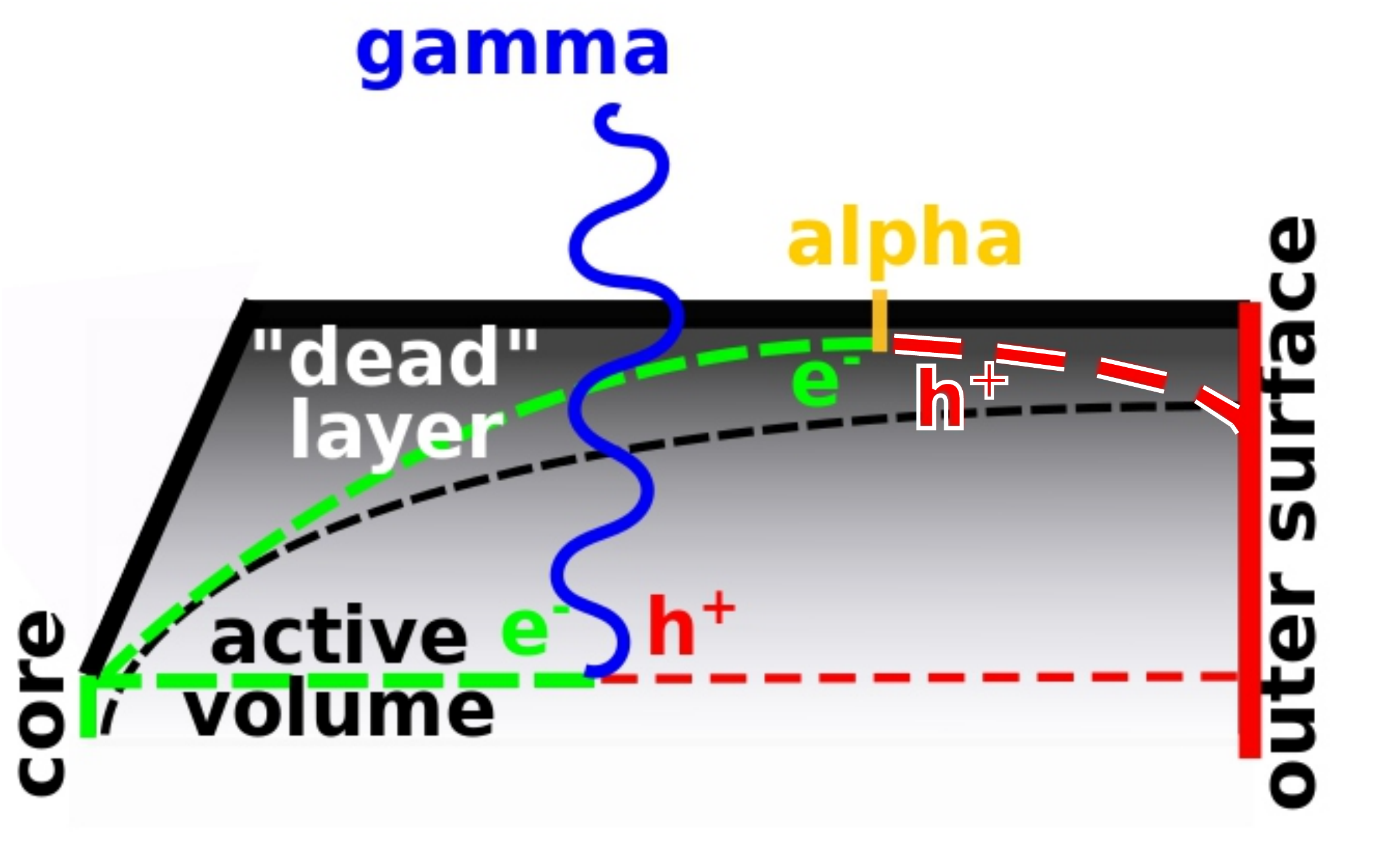}
\caption[Effective dead layer]{Vertical cut through the top of one half of the
Super-Siegfried -- not to scale. The central hole opens conically; the core
electrode does not cover this surface. It is covered 
by a thin passivation layer like the top of the detector. 
The depth and horizontal extension of the conical hole extension
are compatible to the 5~\,mm height of segment~19.
The mantle electrode extends to almost the top of the detector,
the core electrode ends a bit below the conical opening of the hole.
An alpha and a gamma interaction are indicated. An effective dead layer forms 
underneath the passivation. At larger radii, alpha interactions are observed,
even though the drift paths are not straight for charge carriers created in 
this region, adapted from~\cite{Garbini:2016}}
\label{fig:effective_dead_layer}
\end{figure}

The data presented in this work were taken with a 10\,kBq \ce{^{241}Am} 
source, placed inside the collimator above the top end-plate.
The radius of the beam spot was 1.8\,mm. 
Also present was a 42\,kBq \ce{^{152}Eu} calibration source,
located in the collimator
on the side of the detector. 
It was moved down as far as possible  
in order to minimise  its influence on the measurement.

The isotope \ce{^{241}Am} emits alpha particles of 
about $5.5\,\rm MeV$. 
However, the source is encapsulated. 
Separate measurements on the
side of another germanium detector, 
where no significant effective dead layer exists,
showed that the encapsulation reduces the mean
energy of the alphas to 4.5\,MeV with a Gaussian 
variation of 0.2\,MeV~\cite{Garbini:2016}.
The variation corresponds to a 10\,\% inhomogeneity  of the thickness of the
plastic encapsulation.
The energy loss in germanium is 172\,keV/$\mu$m.
Thus, any observation of alpha interactions indicates that the effective 
dead layer is thinner than 26\,$\mu$m.

The \ce{^{241}Am} source also produces gamma-rays of $59.5\,\rm keV$ 
which have a mean free path in germanium of about $0.9\,\rm mm$.
Thus, two different volumes can be probed with 
one source. However, the tungsten collimator also introduces k--$\alpha$
emission lines at 59.3\,keV and 58.0\,keV which cannot be separated
from the 59.5\,keV gammas. They blur the beam spot  
such that the probed
volumes not only differ in depth, but also slightly in radius.

Figure~\ref{fig:Susie} shows the schematic of Super-Siegfried and
the locations of the measurements.
Super-Siegfried is a cylindrical true coaxial n-type high-purity 
segmented germanium detector produced by Canberra France.
It has a height of $70$\,mm. 
The radius of the inner bore hole is $5.05$\,mm and 
the outer radius is $37.5$\,mm.  
According to the manufacturer, 
the detector has a concentration of electrically active impurities 
between \mbox{$\rho = 0.44 \times  10^{10}\textnormal{cm}^{-3}$} 
at the top and \mbox{$\rho = 1.30 \times   10^{10}\textnormal{cm}^{-3}$} 
at the bottom of the detector.  
The full depletion voltage was determined to be 2250\,V. 
The operational voltage was 3000\,V.

The main part of the detector is segmented 6-fold in $\phi$ and 
3-fold in $z$. Such a detector was characterised in detail 
previously~\cite{Abt:2007rf}. 
An additional segment, 5\,mm high and not segmented in $\phi$, 
is located above the regular 18 segments. 
The signals from the detector were recorded as charge pulses.
Charge carriers, electrons and holes created by an energy deposition,
drift towards the electrodes. The electrons drift towards the core electrode
and the holes towards the segment electrodes. During the drift they induce
charges which are recorded. The amount of charge registered 
in a given electrode at a given time
depends on the amount of charge carriers drifting and the strength 
of the so-called weighting field of the electrode.
The weighting field of the core is strong close to the centre 
of the detector while the weighting field of the segment electrodes is
strong close to the mantle. 

Super-Siegfried was previously investigated using the 122\,keV \ce{^{152}Eu}
line with the detector mounted in a conventional cryostat~\cite{Lenz:2010}.
The thin top segment, called segment~19, was
designed to study events on the end-plate. As it is fairly thin, any
charge drifting towards the segment~19 electrode also
creates strong so-called mirror pulses in the segments underneath, 
which facilitate a detailed pulse-shape analysis.

For each event, the pulses for the core and all 19 segments 
were recorded with a 75\,MHz sampling rate. 
All pulses were corrected offline 
for linear cross-talk and calibrated for energy.
The linear cross-talk correction of pulses was performed by
using overall cross-talk factors determined on ADC level from
single segment events~\cite{Garbini:2016}.
For the core and the collecting segments,
the energy was determined after calibration as
the height of the respective pulse using an asymmetric trapezoidal filter. 
Throughout the paper, the risetime of a pulse 
refers to the time for the pulse to develop from 10\% to 90\% of 
its pulse height.

The metalisation of the detector was reduced in order to study
its influence  on the properties of the 
detector\,\footnote{As aluminium is not completely radiopure a reduction
of the metalisation would reduce any background from this source for
low-background experiments~\cite{Majorovits201139}.}. 
Only a circular area with a radius of about $6\,\rm mm$ in the middle 
of each regular segment was metalised, see Fig.~\ref{fig:susieOpenFrame}\,.
The metal contact of segment~19 
only covered a sector of the mantle about 2\,cm long.

Figure~\ref{fig:effective_dead_layer} depicts a schematic vertical cut through
one half of segment~19. 
The conical shape of the bore hole and the limited extension
of the core electrode by construction cause a distorted field
in segment~19, i.e.\ the field lines cannot be horizontal.
In this configuration, a not-fully depleted volume 
with a very weak field close to the hole is expected
for purely geometrical reasons. 
It seems likely that charge carriers are
unable to emerge from this volume.

An effect beyond simple geometry
is also distorting fields in germanium detectors.
In addition, space charges 
accumulating in the thin passivation layer and possibly 
oppositely charged space charges underneath
can distort the field. 
If the field lines are bent sufficiently towards the surface, 
see e.g.\ \cite{dinger:1975}, a  so-called ``surface channel''
forms.
The charge carriers can drift along these channels
with significantly reduced drift speed compared
to the drift speed in the unaffected bulk.
Some studies using either planar detectors or assuming simpler
geometries attribute all observations to surface channel 
effects, see e.g.\ \cite{maggi:2015, utso:2005}. 
However, whether the field is distorted by geometry or by a
space charges, the signature is a reduced speed of the charge carriers. 
The data presented in this paper cannot be used to distinguish 
between the causes.

The partial metalisation of segment~19, see Fig.~\ref{fig:susieOpenFrame}\,,
adds a further complication.
It has the potential to create a phi dependence of the field, because
the Boron implant is not expected to be as conductive as a standard
metalisation layer. The potential on the outer layer is not as
fixed as on a fully metalised detector.

The loss of observed charge can be due to charge carriers created
in a volume from which they cannot emerge or due to charge trapping
along their path to the electrode.
The electrons drift towards the core through a region 
in which the weighting field
of the core is much stronger than for the segment (mantle) electrode.
For the holes, the situation is reversed. Their path towards the segment
electrode is characterised by a strong segment weighting-field.
Thus, the trapping of electrons (holes) affects predominantly 
the energy registered in the core (segment).

Previous measurements with $^{152}$Eu~\cite{Lenz:2010} indicated that
the effective dead layer of the detector close to the bore hole is as thick
as 5\,mm. These measurements were not precise enough measure the thickness
of the effective dead layer at larger radii, even though they indicated 
that the layer was rather thin.
Nevertheless, the expectations~\cite{private:canberra} for this detector 
were that even at larger radii, the thickness of the effective dead layer
would minimally be of the order of 50 to $100\, \rm \mu m$. 
A somewhat similar %highly segmented 
large-volume closed-end detector 
%with a similar end-plate design, 
was probed with 59\,keV photons
and shown to have an effective dead layer
on the scale of millimeters also at larger radii~\cite{eberth:2008}.  
Thus, the observation of alpha particles on the end-plate
with a range of less than 30\,$\mu$m was not expected.

\section{Observation of alpha radiation}
\label{sec:alpha_rad_susie}

\begin{figure}[!h]
\centering
\begin{subfigure}[b]{0.494\textwidth}
  \includegraphics[width=1\linewidth]{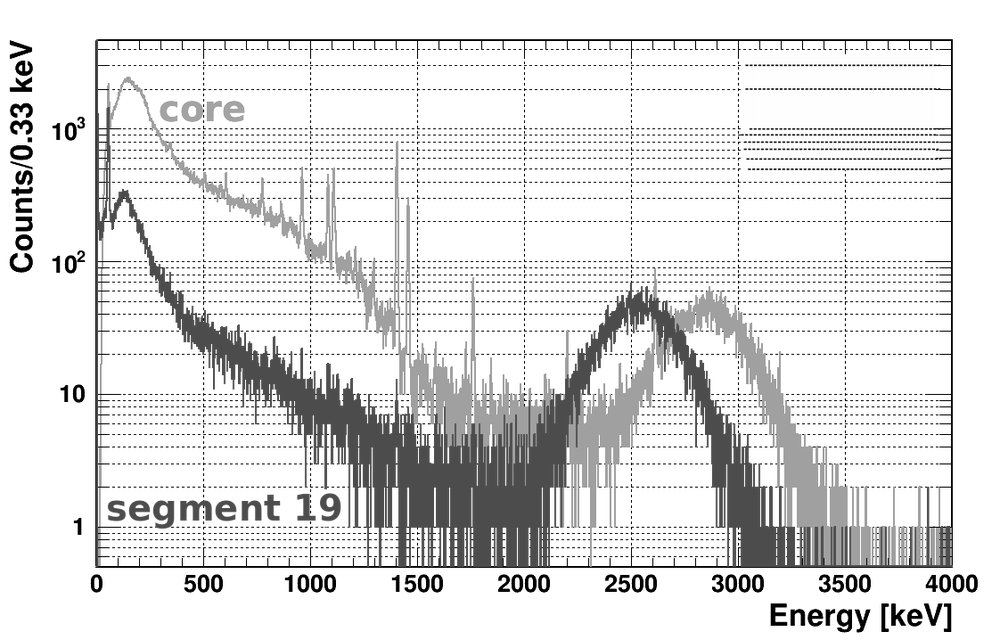}
  \caption[Energy spectra from the core and from segment~19]{}
  \label{fig:alpha_spectrum}
\end{subfigure}
\begin{subfigure}[b]{0.487\textwidth}
  \includegraphics[width=1\linewidth]{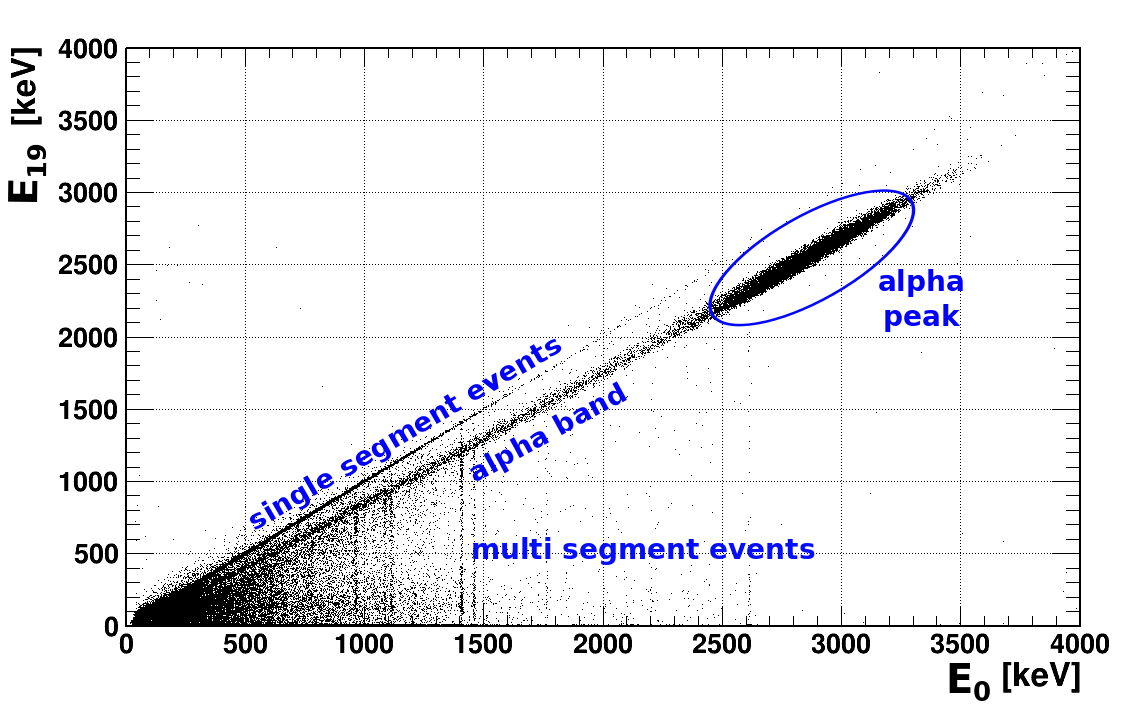}
  \caption[Correlation between the energy recorded in the core 
           and in the top segment]{}
   \label{fig:corrSeg19Core_r26}
\end{subfigure}
\caption[Evidence of alpha interaction in SuSie]
 {Observation of alpha interactions during the scan AS01, 
  see Fig.~\ref{fig:scanning_points}\,, 
  at r$=26.0\,\rm mm$ and $\phi=118^{\circ}$:
  (a) spectra recorded for the core and segment~19,
  (b) correlation between core and segment~19 energies, adapted from~\cite{Garbini:2016}
\label{fig:alpha:ob}
}
\end{figure}

Contrary to expectations, alpha interactions were consistently observed
for radii larger than 25\,mm.
Figure~\ref{fig:alpha_spectrum} shows the spectra recorded by the core 
and by segment~19 for $r=26.0\,\rm mm$ and $\phi=118^{\circ}$, a point covered
by the AS01 scan, see Fig.~\ref{fig:scanning_points}\,. The correlation
between the energies recorded in the core and in segment~19 is shown
in Fig.~\ref{fig:corrSeg19Core_r26}\,.

Broad bumps, from now on referred to as alpha peaks, are dominating
both the core and segment~19 spectra at energies above 2\,MeV.
The core spectrum also shows the characteristic  gamma lines due to natural 
radioactivity  and due to the \ce{^{152}Eu} source. 
Segment~19 is too thin to fully absorb high-energy gammas with
good efficiency, but the correlation plot,
Fig.~\ref{fig:corrSeg19Core_r26}\,, shows a thin line along
the diagonal caused by single-segment~19 events.
The 59~\,keV line is very clearly seen in both the core
and segment~19 and its equal location confirms the validity of
the energy calibration.
Thus, the difference of the location of the alpha peaks
has to be due to a different charge collection 
efficiency for electrons and holes. 
This indicates that holes must be trapped 
to explain the reduced energy observed on segment~19 as
depicted in Fig.~\ref{fig:alpha_spectrum}\,.

The band in Fig.~\ref{fig:corrSeg19Core_r26} associated with alpha events
extends down to extremely low energies. 
However, at low energies, multi-segment events form a background
to the alpha events.
At the highest energies, where the band represents the distinct alpha peaks
of Fig.~\ref{fig:alpha_spectrum}\,, basically no gamma background is present.
High-energy alpha events were selected for further investigation
by requiring events with core (segment~19) energies,
$E_{0}$ ($E_{19}$) of

\begin{equation}
\begin{array}{r@{}l}
    E_{min}(\alpha_{0}^{peak}) < E_{0} < E_{max}(\alpha_{0}^{peak})\, , \\
E_{min}(\alpha_{19}^{peak}) < E_{19} < E_{max}(\alpha_{19}^{peak})\, ,
\end{array}
\label{eqn:alpha_selCriteria}
\end{equation}

where $E_{min}(\alpha_{0 (19)}^{peak})$ and $E_{max}(\alpha_{0 (19)}^{peak})$ are 
defined as the three standard deviation differences from the mean as determined
in respective Gaussian fits.
The background from multi-site gamma events 
was measured to be much less than one percent for this selection
and not considered further. 

\section{Charge trapping}\label{sec:PSA}

\begin{figure}[!h]
 \centering
  \includegraphics[width=1\linewidth]{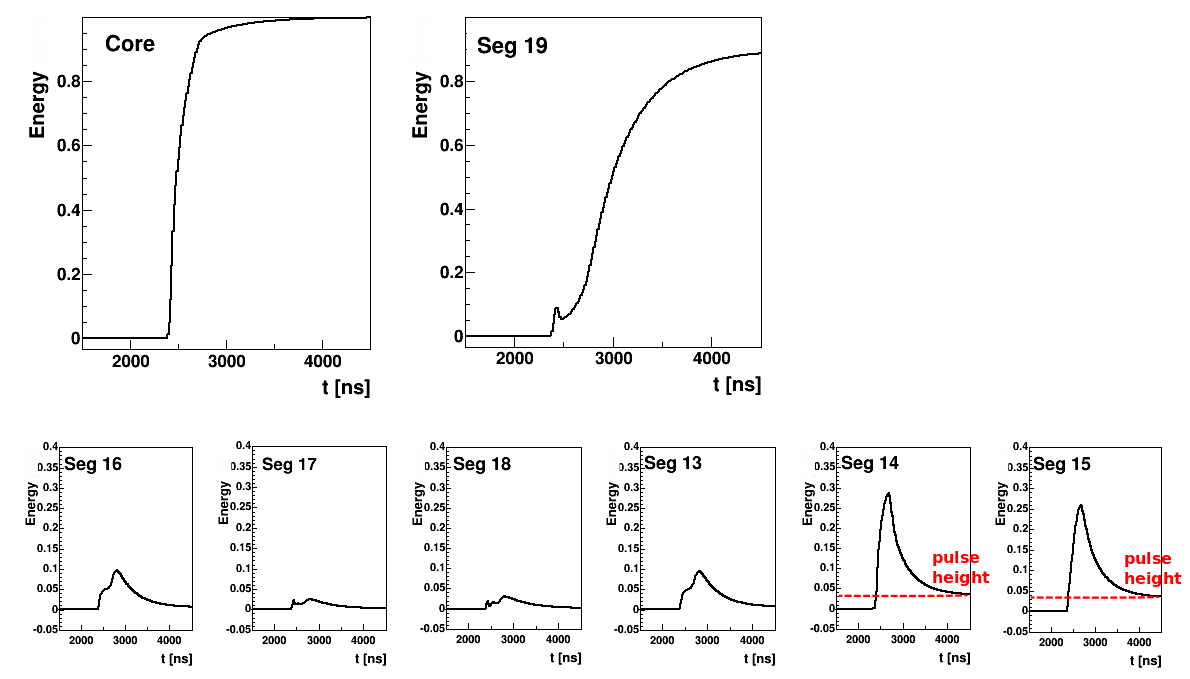}
  \caption[Example of alpha event]{The pulses from the core (top left), 
      segment 19 (top right) and the 6 regular segments below 
      of a typical alpha event 
      from the measurement at $r=26.0\,\rm mm$ 
      and $\phi=118^{\circ}$, in the AS01 set, close to the boundary between 
      segments~14 and~15, adapted from~\cite{Garbini:2016}.
      All pulses are corrected for linear cross-talk.
      The core pulse was normalised to one; all other pulses were 
      normalised to the core pulse, but are shown on a scale up to 0.2.
      The dashed lines in the segment~14 and~15 panels indicate the
      level at which the pulses remain above the baseline.}
  \label{fig:alpha_event}
\end{figure}

Figure~\ref{fig:alpha_event} depicts a typical event 
selected according to Eq.~\ref{eqn:alpha_selCriteria} 
from the measurement at r$=26$\,mm and $\phi=$118\,$^{\circ}$ in the AS01 set.
The pulses from the core (top left), from segment~19 (top right) 
and from the 6 regular segments below are shown in the same time range 
and on the same energy scale. The lower segments only produced extremely
small pulses. 

The core\footnote{The pulse was inverted for easier visual comparison.}
and segment~19 pulses look  as expected for collecting electrodes.
The segment~19 pulse does not reach the height of the core pulse and it
is significantly slower\footnote{The small extra peak at the start of 
the pulse is an artefact created by the cross-talk correction which
does not take differential cross-talk from the core into
segment~19 into account. Such artefacts do not influence any of the
results in this paper.
Differential cross-talk occurs 
due to the very different risetimes of the core and segment~19 pulses.}. 
The core pulse shows a clear kink around 2700\,ns, indicating
the end of the collection of electrons. The slow rise afterwards
is caused by the slow drift of holes towards the mantle.
All segments in the upper layer show positive mirror pulses.
As these segments do not collect any charge, the pulses should
return to the baseline. This happens for segments~13, 16, 17 and 18.
The pulses from segments~14 and 15, directly underneath
the interaction point, however, stay above the baseline;
they are so-called truncated mirror pulses, which were
also described for radiation-damaged detectors~\cite{Descovic:2005}.  

\begin{figure}[!h]
 \centering
  \includegraphics[width=0.75\linewidth]{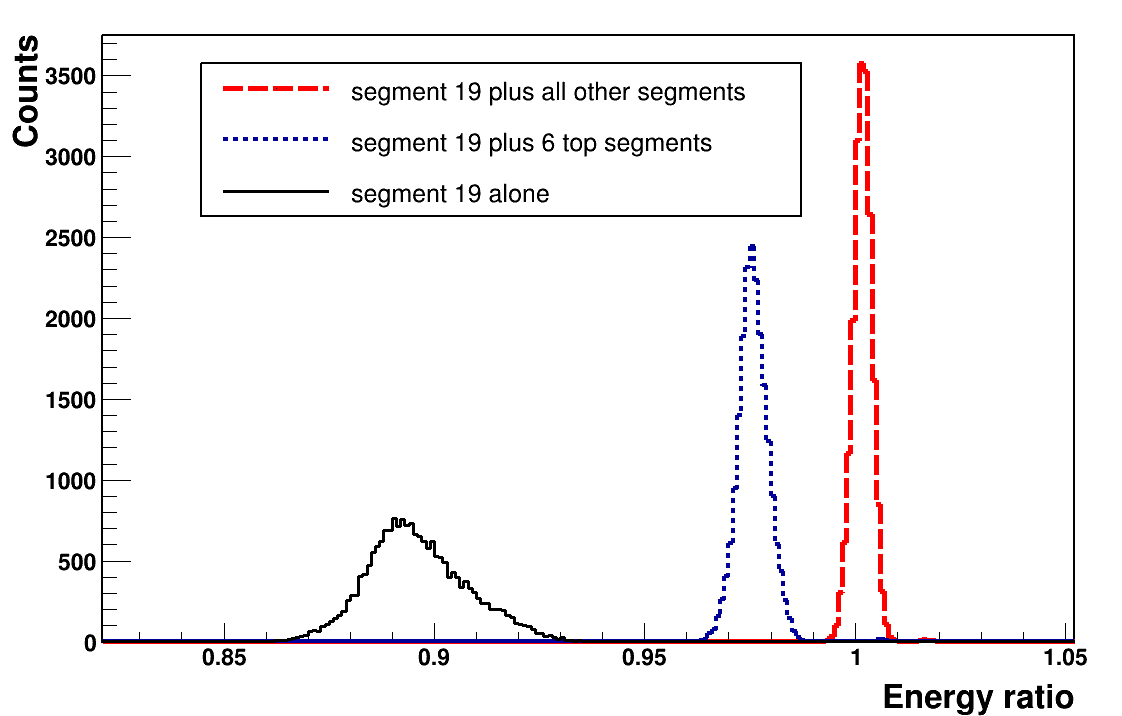}
  \caption[Energy Recovery]{
      Distributions of the ratios $E_{19}/E_0$ (solid line), $(E_{19}+\sum_{i=13}^{18}PH_i)/E_0$ (dotted line) and $(E_{19}+\sum_{i=1}^{18}PH_i)/E_0$ %\footnote{The ratios are defined as: 
%\begin{eqnarray}
%R_{19} = \frac{E_{19}}{E_0} \, , \\
%R_{19+UP} = \frac{E_{19}+\sum_{i=13}^{18} PH_i}{E_0} \, ,\\
%R_{19+ALL} = \frac{E_{19}+\sum_{i=1}^{18} PH_i}{E_0} \,.
%\end{eqnarray}      
%      } 
(dashed line) for the alpha events      
      from the measurement at $r=26.0\,\rm mm$ 
      and $\phi=118^{\circ}$, in the AS01 set, close to the boundary between 
      segments~14 and~15.
      }
  \label{fig:energy-recovery}
\end{figure}

Truncated mirror pulses are the signature of charge trapping.
The amount of charge-trapping can be determined as the
final height above the baseline at which a mirror pulse remains.
This we define as the pulse-height, $PH_i$, of a non-collecting segment $i$.
Adding these pulse-heights of the non-collecting segments 
to the energy recorded for segment~19, 
the energy corresponding to the core pulse can be recovered.
This is demonstrated in Fig.~\ref{fig:energy-recovery}
The ``energy recovery'' works to a level better than 0.3\,\% when
all segments are considered.
A Gaussian fit to the rightmost distribution in Fig.~\ref{fig:energy-recovery}
yields a mean of 1.002 and a width of 0.25\,\%, corresponding to less than
10\,keV. 
Figure~\ref{fig:energy-recovery} also shows that
though the mirror pulses in the lower segments are very small, they
are needed to completely balance the energy. 

For this source position at
$r=26.0\,\rm mm$ and $\phi=118^{\circ}$ in the AS01 set,
100\,\% of the selected alpha-induced events contain such truncated 
positive mirror pulses; the pulses were very reproducible~\footnote{The
somewhat unexpected reproducibility of the alpha event pulses was 
observed for all locations.}.
The segment~19 pulse has not completely flattened out at the end
of the recorded time window. It is possible that some holes
were released later. The time window was unfortunately limited by 
the data acquisition system. 

The absence of negative truncated mirror pulses 
at this position does not exclude
that also electrons are trapped. As the holes drift outwards,
they come closer to the segment electrodes than the electrons and
thus the large positive charge they induce makes the observation of
a smaller amount of induced negative charge indicating
electron trapping impossible.

The long pulse shown for segment~19
in Fig.~\ref{fig:alpha_event} shows that the 
holes drift at a reduced speed. 
That makes them more likely to get trapped along their way to the mantle.

\begin{figure}[]
 \centering
   \includegraphics[width=1\linewidth]{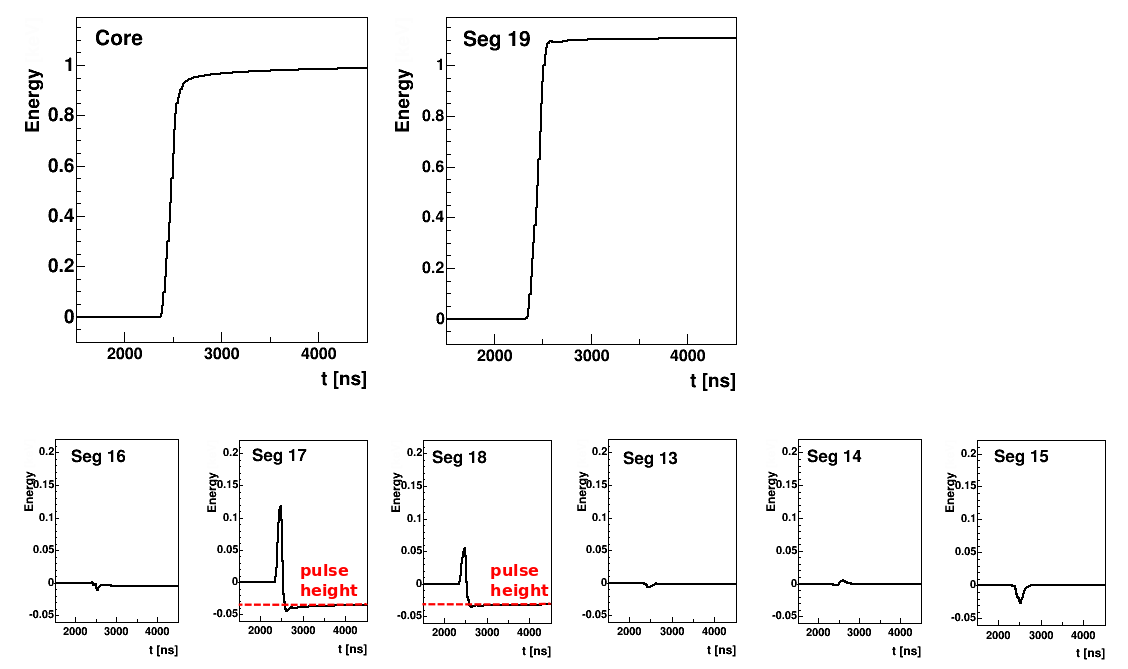}
   \caption[Example of a fast alpha event]{The pulses from the core (top left), 
      segment~19 (top right) and the 6 regular segments below 
      of a typical alpha event from the measurement at $r=38.0\,\rm mm$ 
      and $\phi=312^{\circ}$, in the RS02 set, 
      close to the boundary between segments~17
      and~18, adapted from~\cite{Garbini:2016}.
      All pulses are corrected for linear cross-talk.
      The core pulse was normalised to one; all other pulses were 
      normalised to the core pulse, but are shown on a scale up to 0.2. 
%%%%%      The mirror pulses observed in segments~17 and~18 are shown
%      enlarged in Fig~\ref{fig:fast_event_segments}.
}
   \label{fig:fast_event}
 \end{figure}

%\begin{figure}[]
% \centering
% \begin{subfigure}[b]{0.4\textwidth}
%   \includegraphics[width=1\linewidth]{eventFast_seg17.png}
%   \caption[Fast event in segment 17]{}\label{fig:fast_event_17}
% \end{subfigure}
% \begin{subfigure}[b]{0.4\textwidth}
%   \includegraphics[width=1\linewidth]{eventFast_seg18.png}
%   \caption[Fast event in segment 18]{}\label{fig:fast_event_18}
% \end{subfigure}
%   \caption[Segments of the fast alpha event]
%      {Enlargement of the pulses from the event 
%       in Fig~\ref{fig:fast_event}: 
%       (a) segment~17,
%       (b) segment~18.
%}
%  \label{fig:fast_event_segments}
%\end{figure}

At larger radii, also electron trapping was observed.
Figure~\ref{fig:fast_event} shows an event at $r = 38.0$\,mm in the 
RS02. This event is located at the edge of the detector and
close to the sector of metalisation.  
Figure~\ref{fig:fast_event} top right shows 
that the energy collected
in segment~19 is larger than in the core and that
the drift of the holes
is fast in this region.
The mirror pulses in segments~17 and~18 
%shown in 
%Figs.~\ref{fig:fast_event_17} and ~\ref{fig:fast_event_18} 
first reflect the drift of
the positive holes and then the drift of the electrons. The
electrons are trapped
at $t \approx 2600$\,ns. 
The other mirror pulses are much smaller. 
A very slow release of electrons is
indicated by the slow rising tails of the mirror pulses 
and the steady increase of the core pulse at $t > 2600$\,ns.

\section{Location dependence of charge trapping}
\label{sec:loc:results}

\begin{figure}[!h]
 \centering
  \includegraphics[width=0.7\linewidth]{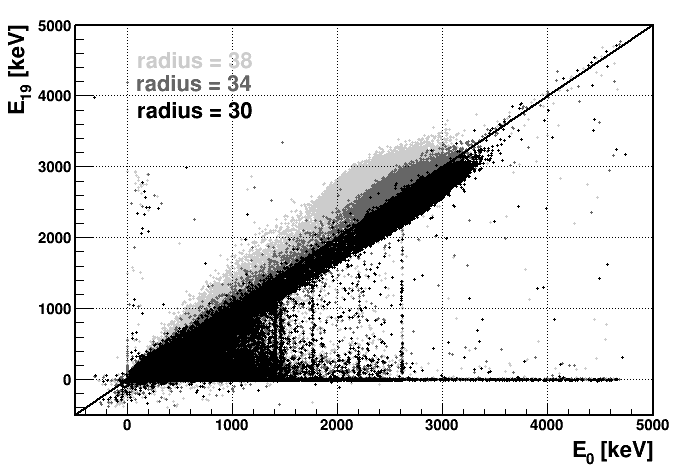}
\caption[Correlation between $E_0$ and $E_{19}$ for different radii]
        {Correlation between $E_0$ and $E_{19}$ for measurements at
         different radii 
         at $\phi=33^{\circ}$, RS01, see Fig.~\ref{fig:scanning_points}\,.
         The line along the diagonal indicates  $E_0 = E_{19}$, adapted from~\cite{Garbini:2016}.} 
\label{fig:E19corrE0}
\end{figure}

Figure~\ref{fig:E19corrE0} shows the correlation between the 
energy in segment~19, $E_{19}$, and the energy in the core, $E_0$,  
for the three different radii probed by the RS01 set 
as introduced in Fig.~\ref{fig:scanning_points}\,.

With increasing radius, the alpha band moves from below the diagonal, 
normally indicating single segment events, to above.
The core registers more energy for small radii and segment~19 registers
more energy for large radii, close to the edge of the detector.
The change in observed core energy suggests that either the really
inactive volume of the detector becomes thicker towards larger radii
or that also electrons get trapped. 
As the energy observed in segment~19 increases, the latter explanation is
more consistent. 
This agrees also with expectations.
As the radius increases, the path of the electrons becomes longer
and they are more likely to be affected by the field distortions.
The situation is reversed for holes. Their path shortens and their trapping
becomes less probable.

In order to systematically study the radial dependence of charge trapping, 
truncated mirror pulses were methodically searched for.
For all positions and selected alpha events, 
the pulses of the segment underneath the interaction zone
were classified~\cite{Garbini:2016}. 
The fractions $f^{\rm pt}_\alpha$ of pulses falling into six
``pulse types''(pt),  ``mirror positive'',  ``mirror truncated positive'', 
``mirror negative'',  ``mirror truncated negative'', 
``normal collecting'' and ``no activity'', were calculated. 
\begin{figure}[!h]
\centering
 \begin{subfigure}[b]{0.428\textwidth}
   \includegraphics[width=1\linewidth]{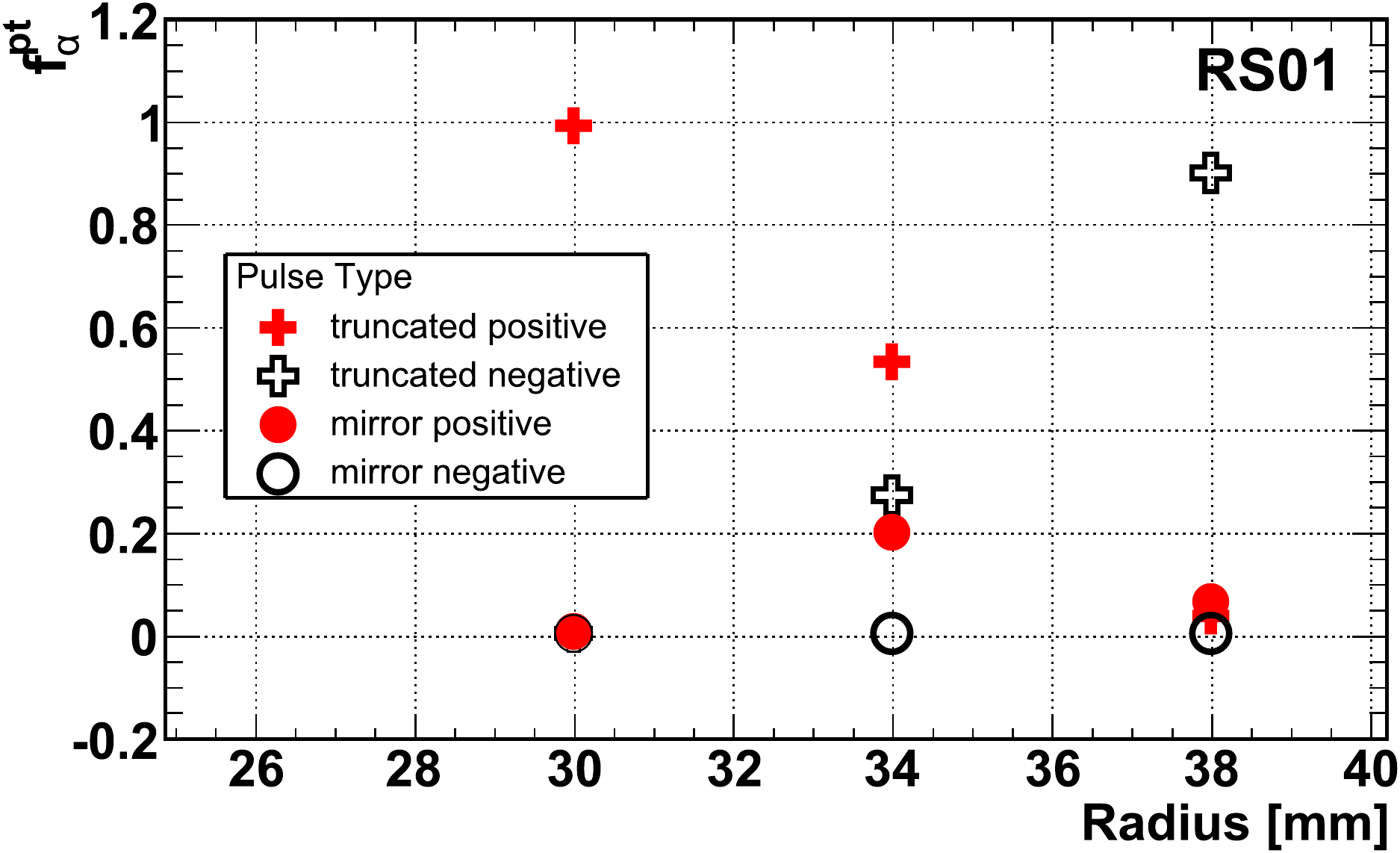}
   \caption[$f^{\rm pt}_{\alpha}$ as function of the radius at $\phi=33^{\circ}$]{}
   \label{fig:alpha_pulseStat_radial32}
 \end{subfigure}%
\begin{subfigure}[b]{0.452\textwidth}
  \includegraphics[width=1\linewidth]{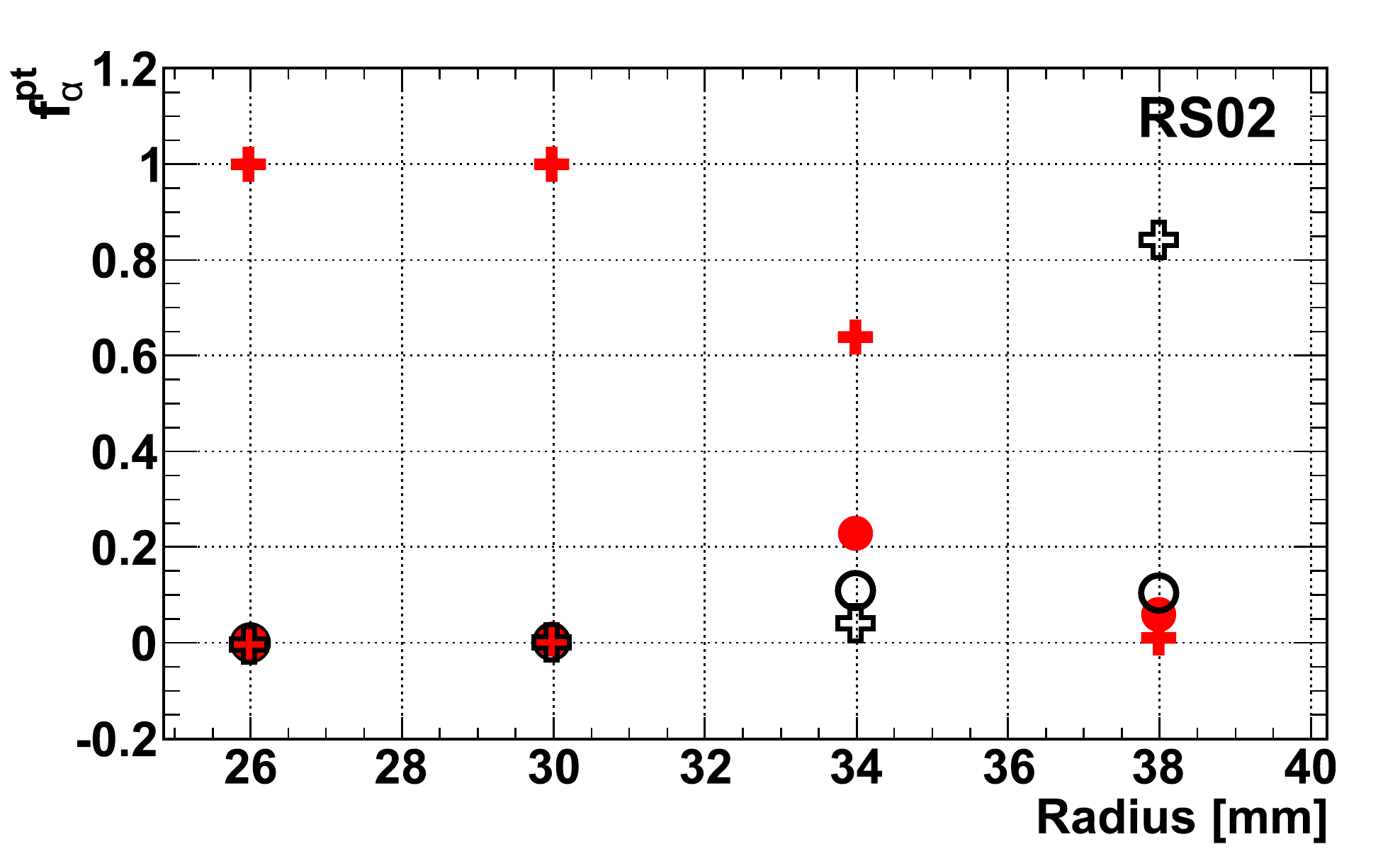}
  \caption[$f^{\rm pt}_{\alpha}$ as function of the radius at $\phi=312^{\circ}$]{}\label{fig:alpha_pulseStat_radial312}
\end{subfigure}
\begin{subfigure}[b]{0.43\textwidth}
  \includegraphics[width=1\linewidth]{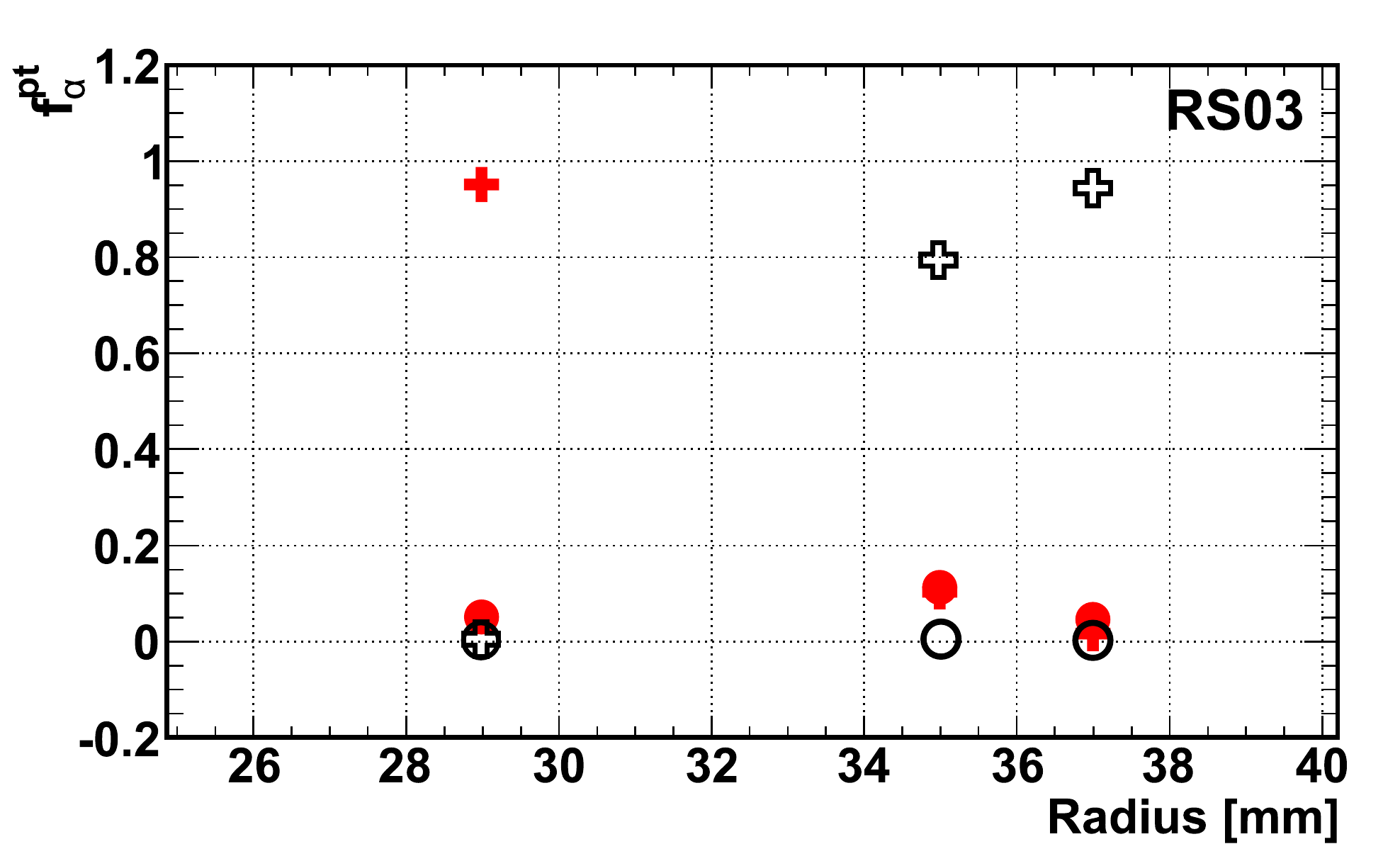}
  \caption[$f^{\rm pt}_{\alpha}$ as function of the radius at $\phi=176^{\circ}$]{}\label{fig:alpha_pulseStat_radial176}
\end{subfigure}
\caption[Pulse statistics for selected alpha events in the radial scans]
        {$f^{\rm pt}_{\alpha}$ as a function of the radius 
         for different azimuthal angles: 
         (a) $\phi=33^{\circ}$,  RS01, 
         (b) $\phi=312^{\circ}$, RS02, 
         (c) $\phi=176^{\circ}$, RS03, adapted from~\cite{Garbini:2016}.
}
\label{fig:alpha_pulseStat_radialScan}
\end{figure} 
Figure~\ref{fig:alpha_pulseStat_radialScan} shows $f^{\rm pt}_\alpha$ 
as a function of the radius for the three different 
azimuthal angles associated with the scans RS01, RS02 and RS03. 
The results are very consistent at the different angles.
Up to $r=30$\,mm, all events show positive truncated mirror-pulses,
indicating hole trapping. At $r=34$\,mm, the fractions of
positive truncated mirror-pulses drops and negative truncated mirror-pulses
become observable. At the edge of the detector, negative truncated
mirror-pulses dominate, indicating electron trapping.
The radial dependence of the effect is the same for RS02 close to the
metalisation as for the other scans. 

The results obtained for the two azimuthal scans AS01 and AS02 
at $r=26$\,mm and  $r=30$\,mm confirm the results of the radial scans.
For both scans, 100\,\% of the events contain positive truncated 
mirror-pulses in the respective segments underneath segment~19.

Truncated mirror pulses were also searched for in gamma events
at the scan positions where alpha events were observed. 
None were found. This confirms that charge trapping is restricted
to charge carriers created very close to the surface.

\begin{figure}[!h]
\centering
\begin{subfigure}[b]{0.49\textwidth}
  \includegraphics[width=1\linewidth]{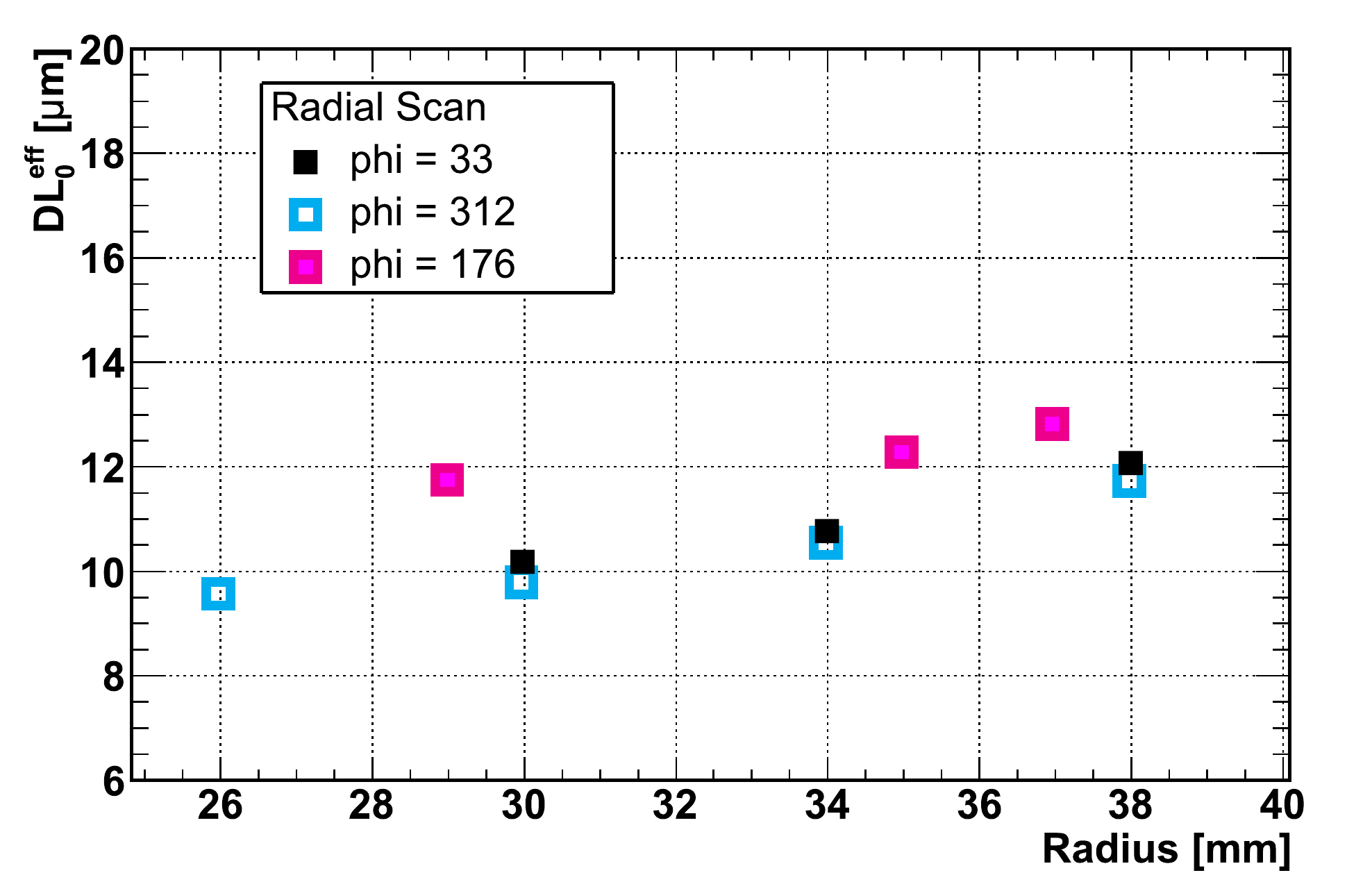}
  \caption[$\rm DL^{\rm eff}_{0}$ as a function of the radius]{}\label{fig:deadLayer_core_radial}
\end{subfigure}
\begin{subfigure}[b]{0.49\textwidth}
  \includegraphics[width=1\linewidth]{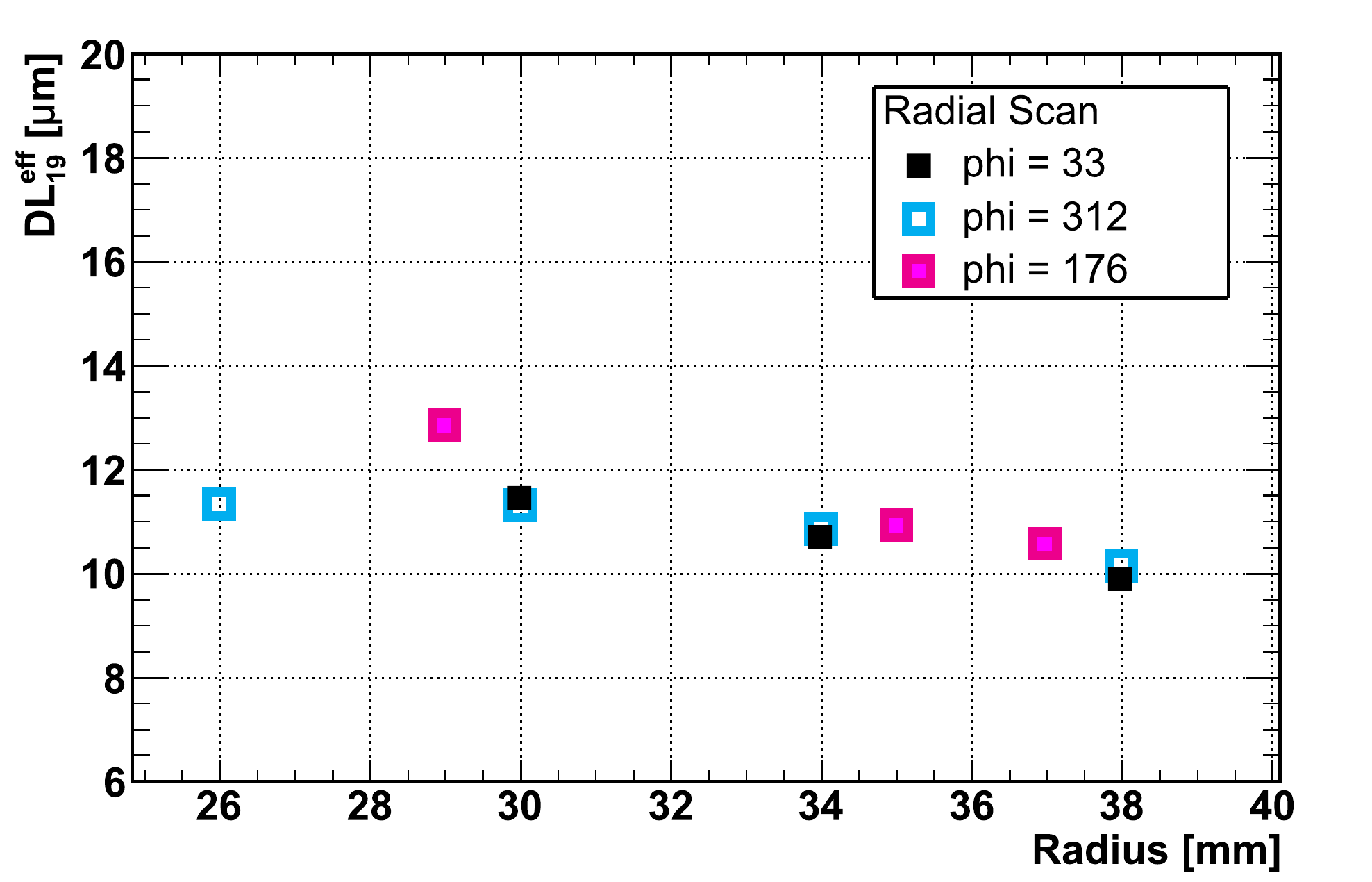}
  \caption[$\rm DL^{\rm eff}_{19}$ as a function of the radius]{}\label{fig:deadLayer_seg19_radial}
\end{subfigure}
\caption[Thickness of the $\rm DL^{\rm eff}$ as a function of the radius]
         {Effective dead layer thicknesses, 
          $\rm DL^{\rm eff}_{i}$, as a function of the radius for 
          (a) the core, $i=0$, and (b) segment~19, $i=19$. 
           The statistical uncertainties as determined by the 
           fitting procedure are smaller than the symbol size, adapted from~\cite{Garbini:2016}.}
\label{fig:eff_deadLayer_radial}
\end{figure}

The observed energies in the core and segment~19 were converted to
``effective thicknesses'',
${DL}^{\textrm {eff}}_i$, of the effective dead layer as 
\begin{equation}
\textrm{DL}^{\textrm {eff}}_i = \left( \frac{d E}{d x}\right) ^{-1}(E^{\rm ini} - \mu_{i})\,~, 
\label{eq:dead_layer}
\end{equation}
where $i=0$ or $19$, $E^{\rm ini}$ is the average incident alpha energy and
$\mu_i$ is the observed average energy for the selected alpha events.
This assumes that the energy observed is related to the energy lost in
an inactive volume of the detector. This is clearly not the case as
energy is also lost due to charge carrier trapping.
However, for an experiment which only records energies, this is a convenient
and common nomenclature.
The results are depicted in Fig.~\ref{fig:eff_deadLayer_radial}\,.

\begin{figure}[!h]
\centering
\begin{subfigure}[b]{0.49\textwidth}
  \includegraphics[width=1\linewidth]{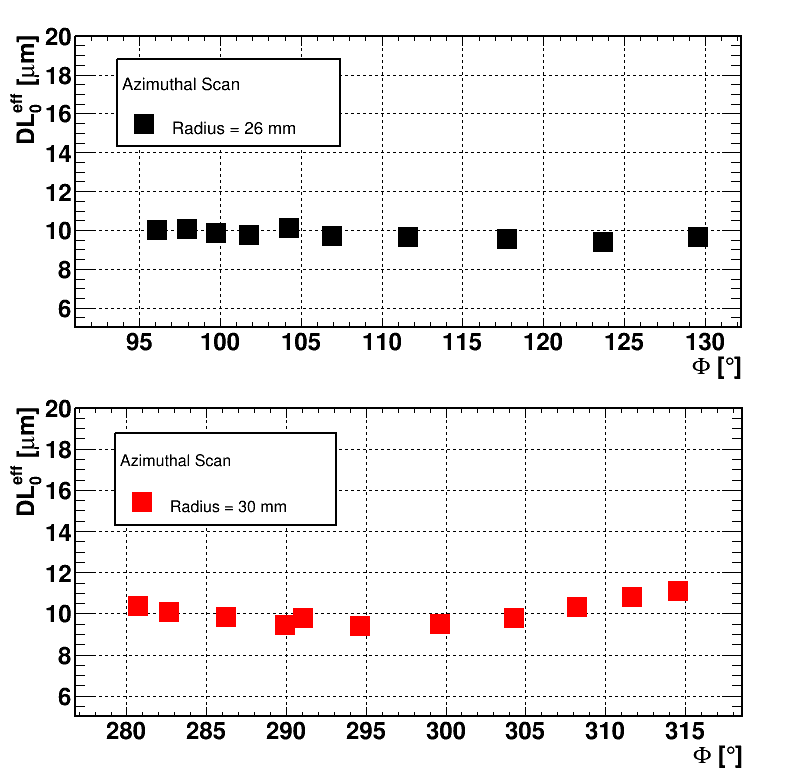}
  \caption[$\rm DL^{\rm eff}_{0}$ as a function of the azimuth]{}\label{fig:deadLayer_core_azimuthal}
\end{subfigure}
\begin{subfigure}[b]{0.48\textwidth}
  \includegraphics[width=1\linewidth]{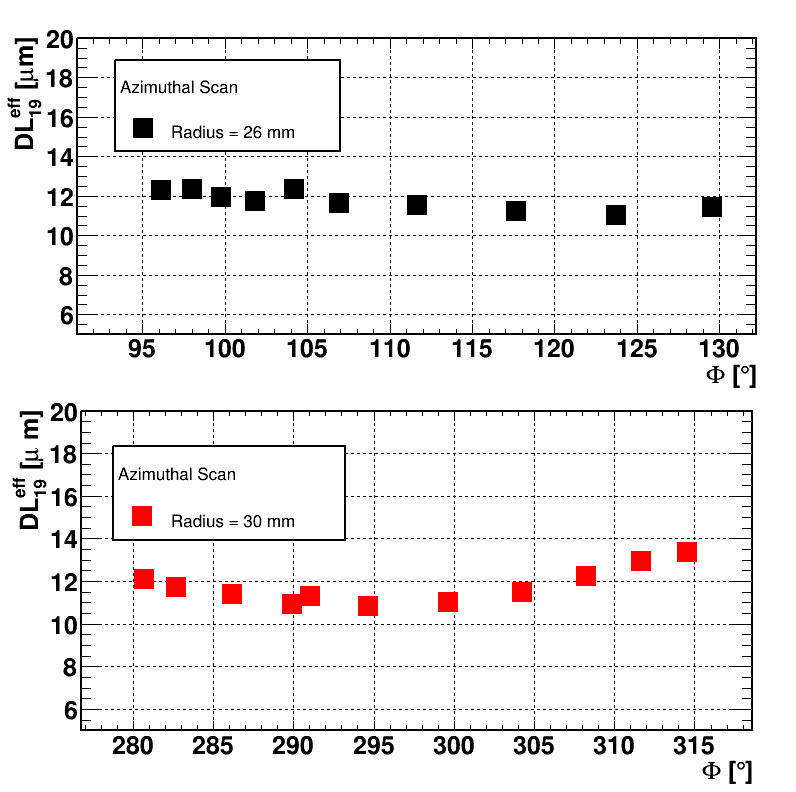}
  \caption[$\rm DL^{\rm eff}_{19}$ as a function of the azimuth]{}\label{fig:deadLayer_seg19_azimuthal}
\end{subfigure}
\caption[Thickness of the $\rm DL^{\rm eff}$ as a function of the azimuth]
         {Effective dead layer thicknesses, 
          $\rm DL^{\rm eff}_{i}$, as a function of the azimuthal angle for 
          (a) the core, $i=0$, and (b) segment~19, $i=19$. 
           The statistical uncertainties as determined by the 
           fitting procedure are smaller than the symbol size, adapted from~\cite{Garbini:2016}.}
\label{fig:eff_deadLayer_azimuthal}
\end{figure}

Corresponding to the decrease of observed energy in the core at increasing
radii, $\rm DL^{\rm eff}_{0}$ slightly increases, 
see Fig.~\ref{fig:deadLayer_core_radial}\,.
The situation for the holes, see Fig.~\ref{fig:deadLayer_seg19_radial}\,,
is reversed. 
The conditions are very similar for all three radial scans.
At $\phi=176^{\circ}$, the values of $\rm DL^{\rm eff}_{0}$ 
and $\rm DL^{\rm eff}_{19}$ at $r=29$\,mm are slightly larger than for
the other locations.
This area is far away from the metalisation.
However, the area covered by the azimuthal scan
AS01 from $\phi=95^{\circ}$ to $\phi=130^{\circ}$
at $r=26$\,mm is equally far away from the metalisation and
does not show thicker effective dead layers.
The values are 
$\rm DL^{\rm eff}_{0} \approx 10\,\mu$m 
and $\rm DL^{\rm eff}_{19} \approx 12\,\mu$m and are quite constant, 
see Fig.~\ref{fig:eff_deadLayer_azimuthal}(top row).

The scan AS02, 
close to RS02 and covering the metalisation at $r=30$\,mm, 
revealed a shallow dependence on $\phi$ of 
both $\rm DL^{\rm eff}_{0}$ and $\rm DL^{\rm eff}_{19}$,
see Fig.~\ref{fig:eff_deadLayer_azimuthal}(bottom row).
At the centre of the metalisation, both drop to about 10\,$\mu$m.
It should be noted again that $\rm DL^{\rm eff}_{0}$ and $\rm DL^{\rm eff}_{19}$
are just a convenient nomenclature. The drop is more likely 
due to less charge trapping.
The smallest value of $\rm DL^{\rm eff}_{0}$
could indicate a truly dead layer. But this is also rather unlikely.
In Figs.~\ref{fig:corrSeg19Core_r26} and~\ref{fig:E19corrE0}\,, events
with observed energies in the core and segment~19 well above the
alpha peaks are visible. Thus, there are events in which the charge carriers
escape and make it all the way to the electrodes.
On the other hand, the bands also extend to very low observed energies.
For these events, almost all charge carriers cannot be collected, even
though they were generated down to 26\,$\mu$m.
The events in the observed alpha peaks are very reproducible.
However, the other events cover a large phase-space of observed energies.

In general, the situation is quite similar at all locations where alpha
interactions were observed.
Alpha-dedicated measurements were also performed at radii 
smaller than $25\,\rm mm$, see Fig.~\ref{fig:scanning_points}\,. 
However, alpha peaks were not observed 
anywhere for these smaller radii. 
The thickness of the inactive volume of the detector
is rapidly growing in this region. This was also previously observed
by probing with low energy gammas~\cite{Lenz:2010}.

\section{Risetimes of pulses}

\begin{figure}[!h]
\centering
\begin{subfigure}[b]{0.49\textwidth}
  \includegraphics[width=1\linewidth]{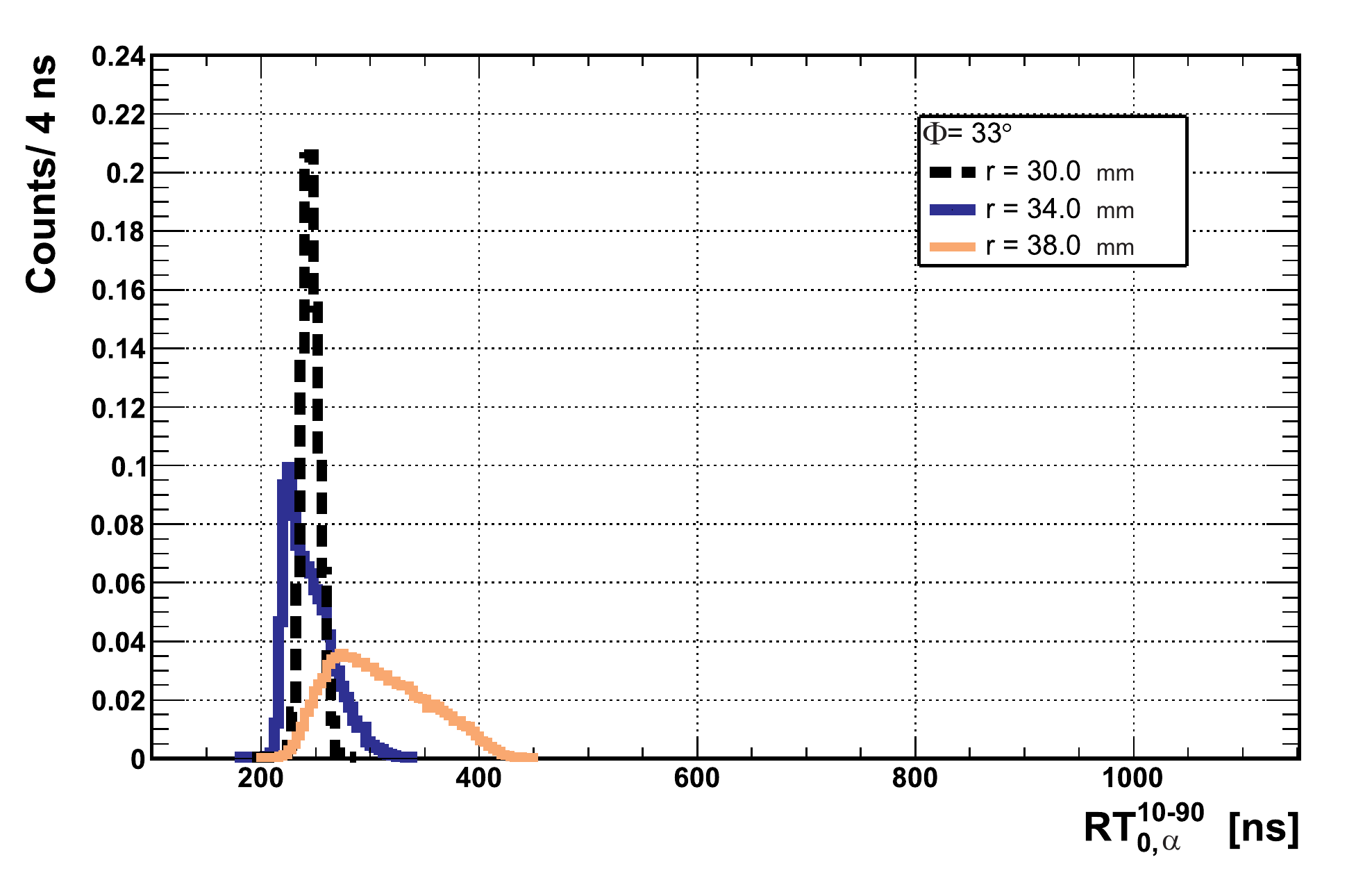}
  \caption[]{}\label{fig:risetime-distr-a}
\end{subfigure}
\begin{subfigure}[b]{0.49\textwidth}
  \includegraphics[width=1\linewidth]{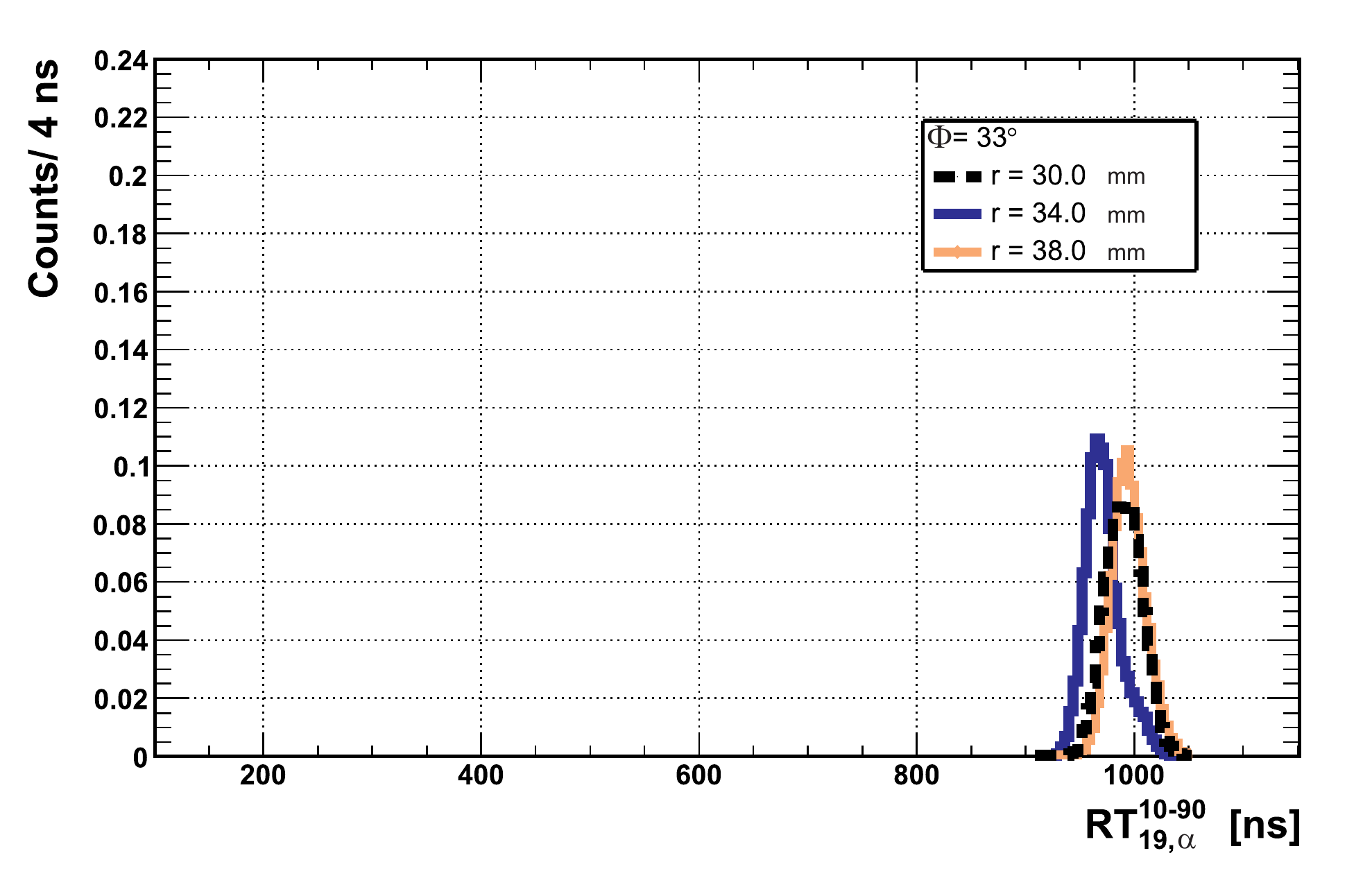}
  \caption[]{}\label{fig:risetime-distr-b}
\end{subfigure}
\caption[Radial dependence of the risetime]
         {Distribution of risetimes 
          (a) $RT_{0}^{10-90}$ for the core and (b) $RT_{19}^{10-90}$ for 
          segment~19 for the radial scan RS01, adapted from~\cite{Garbini:2016}.
}
\label{fig:risetime-distr}
\end{figure}

The risetime of pulses, $RT_i^{10-90}$, presented and discussed here 
always denotes the time in which the pulse rises from 10\,\% to 90\,\% of 
its amplitude.
As the pulses recorded for the core are very different than for segment~19,
this results in very different risetimes.
The distributions of the risetimes for the radial scan RS01 are shown
in Fig.~\ref{fig:risetime-distr}\,.

The pulses recorded in segment~19 rise much slower than the core pulses.
This is possible because the holes 
drifting close to the mantle contribute little to the core pulse 
as the core weighting field is small in this region.  
The pulses in segment~19 are extremely long.
The length of the expected drift path does not seem to be the
determining fact, but rather the passage through a volume where the electric field lines are extremely distorted and the holes drift slowly.
The situation is different for the core pulses.
They get longer for larger radii in accordance with the longer drift path.

\begin{figure}[!h]
\centering
\begin{subfigure}[b]{0.49\textwidth}
  \includegraphics[width=1\linewidth]{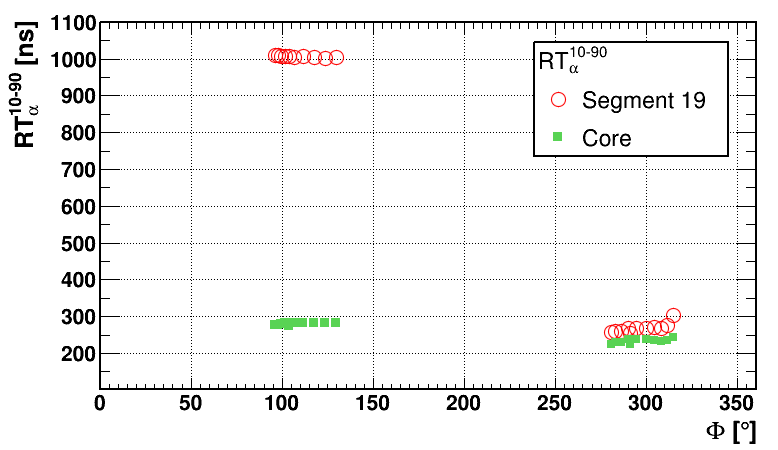}
  \caption[]{}\label{fig:rtAllRange_alpha}
\end{subfigure}
\begin{subfigure}[b]{0.49\textwidth}
  \includegraphics[width=1\linewidth]{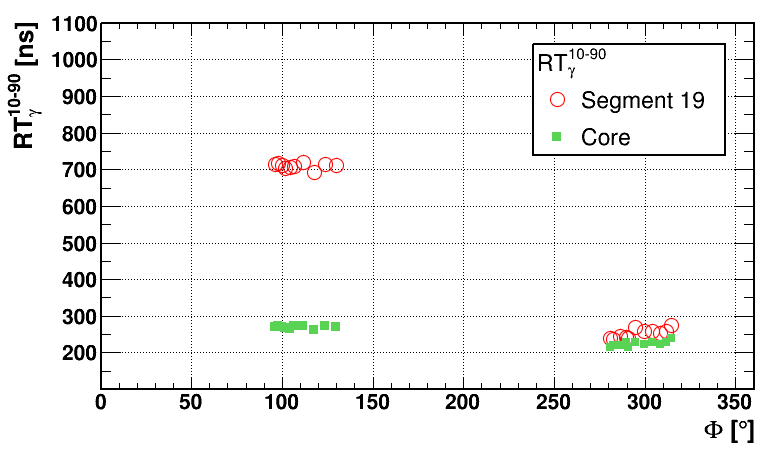}
  \caption[]{}\label{fig:rtAllRange_gamma}
\end{subfigure}
\caption[Azimuthal dependence of the risetime]
 {Azimuthal dependence of the average $RT^{10-90}$ for the core (square) 
  and segment~19 (open circle) for (a) selected alpha events,
  (b) $59.5\,\rm keV$ gamma events.}
\end{figure}

The mean values of the 
risetime distributions of alpha events are shown for 
the two azimuthal scans in Fig.~\ref{fig:rtAllRange_alpha}\,. 
The core pulses have average risetimes of slightly less than
300\,ns for both scans. The average risetimes of the segment~19
pulses are however very different for the two scans.
In the sector around the metalisation,
the segment~19 pulses rise almost as fast as the core pulses.
This indicates that the electric field is significantly less distorted
in the sector with the metalisation than in the other sectors.

Figure~\ref{fig:rtAllRange_gamma} shows the average risetimes
for $59.5\,\rm keV$ gamma events.  
These gammas almost probe the complete volume of the top segment. 
Some gammas interact close to critical areas of the detector, where the 
charge collection is extremely slowed down. Some gammas, however, 
interact far away from such critical area. This causes extremely 
broad risetime distributions. The mean of these distributions lies
significantly above the mean of the core distributions for the sector
far away from the metalisation. This indicates that the distortion
of the field is not restricted to the very thin layer probed by alpha
particles. 

This is confirmed by a side-scan of segment~19 which was performed
earlier~\cite{Lenz:2010} in the K1~\cite{Abt:2007rf} vacuum cryostat. 
A $42\,\rm kBq$ collimated \ce{^{152}Eu} source was rotated around the
detector at $z = 67.5$\,mm, the middle of segment~19 (see~\ref{fig:susieOpenFrame}\,), 
in steps of about 10$^{\circ}$. 
Single-segment~19 events with an energy of $121.78\pm 5\,\rm keV$, corresponding to a strong gamma line of \ce{^{152}Eu}, were selected.   
The mean free path of such gammas is about 4\,mm
and the dominant interaction cross section 
is associated to the photo effect. Thus, most interactions create a single
energy deposit close to the outer mantle. 

\begin{figure}[!h]
\centering
\begin{subfigure}[b]{0.49\textwidth}
  \includegraphics[width=1\linewidth]{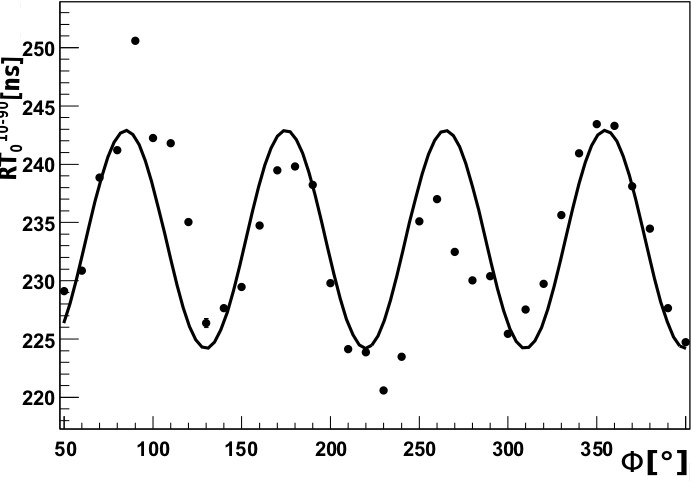}
  \caption[]{}\label{fig:risetime_core_daniel}
\end{subfigure}
\begin{subfigure}[b]{0.49\textwidth}
  \includegraphics[width=1\linewidth]{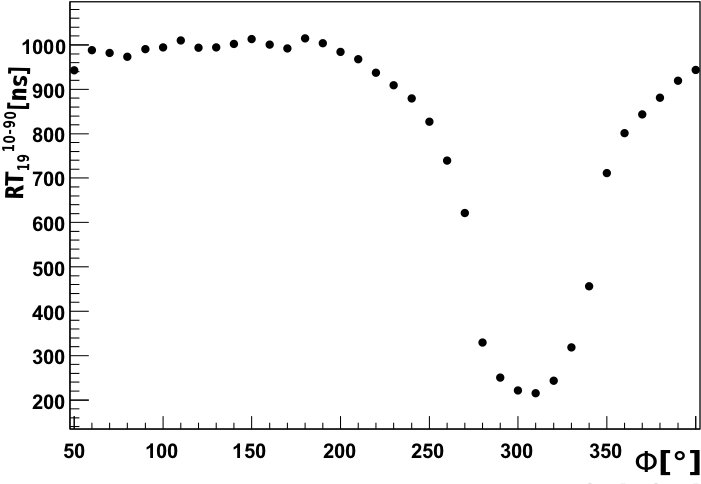}
  \caption[]{}\label{fig:risetime_seg19_daniel}
\end{subfigure}
\caption[]{The mean $RT_i^{10-90}$ (a) for the core, $i=0$, and (b) for
    segment~19, $i=19$ as a function of $\phi$ as determined from 
    a side-scan of the detector, adapted from~\cite{Lenz:2010}.}
\label{fig:1090CoreSeg}
\end{figure}

Figure~\ref{fig:1090CoreSeg} shows the mean risetimes for (a) the core
and (b) segment~19.
Again, the difference is striking. 
While the core pulses show the normal small variation 
in risetime depending on their relative location
to the crystal axes, the segment~19 pulses only show ``normal'' risetimes
close to the  metalisation.  
For locations far away from the metalisation, the risetimes are as large as
for the alpha induced events. 
This proves that the effect of the missing metalisation is not confined to
the very top of segment~19.

\section{Conclusions}\label{sec:conclusions}

The observation of alpha events over most of the surface of the 
passivated end-plate of a true-coaxial detector was not really expected.
It demonstrates that the effective dead layer underneath the passivation layer
is extremely thin, only of the order of 10\,$\mu$m thick.
The spectra resulting for alpha interactions in the
end-plate cover the region around 2\,MeV where the signal for neutrinoless 
double beta decay would appear. If only the energy registered in core
is recorded, such an alpha event could be mistaken for a signal event.
A simple analysis of the shape of the core pulse would not help, because
its risetime is normal.

In order to identify such events on the surface of an end-plate,
it is necessary to read out the mantle.
The difference between the energies recorded for the core and the mantle
can tag alpha induced surface events.

Surface events are characterised by charge trapping, both of holes
and electrons. However, to clearly identify these trapping effects,
it is necessary to have the mantle segmented. This might not be
favourable for a large scale experiment with hundreds of detectors.
A possible handle might be a special configuration of the metalisation
close to the end-plate.
Contrary to intuition, it might be favourable to leave a small ring
of the detector close to the end-plate without metalisation.
A ring height of 5\,mm is definitely
too much because it affects an equally deep layer, but a ring height
matching the volume to be tagged could help to identify events close to the
end-plate through the slow rise of their mantle pulses.

\section*{Acknowledgements}
We would like to thank the staff of Canberra France, now Mirion technology,
for their cooperation.

%\bibliography{<your-bib-database>}
%\section*{References}
\bibliographystyle{elsarticle-num}
\bibliography{mybibfile}

\end{document}